\documentclass[aps,prd,twocolumn,superscriptaddress]{revtex4}
 \usepackage[english]{babel} 
 \usepackage{latexsym} 
 \usepackage{amsmath} 
 \usepackage{amsfonts} 
 \usepackage{amssymb} 
 \usepackage{textcomp} 
 \usepackage{multirow}
 \usepackage{graphicx}  
 \usepackage{enumerate} 
 \usepackage{array} 
 \usepackage{tikz}
 \usetikzlibrary{fit,calc,positioning,decorations.pathreplacing,matrix}
 \usepackage[latin3]{inputenc} 
 \usepackage{graphics} 
 \usepackage{color} 
\usepackage{hyperref}
 
 \newcommand{\lra}{\leftrightarrow}
  
 \newcommand{\be}{\begin{equation}} 
 \newcommand{\ee}{\end{equation}} 
 \newcommand{\ba}{\begin{eqnarray}} 
 \newcommand{\ea}{\end{eqnarray}} 

\def\bs{\begin{subequations}}
	\def\es{\end{subequations}}

\def\g{\gamma}

\def\n{\nu}
\def\m{\mu}

\begin{document} 
%opening 
%%%%%%%%%%%%%%%%%%%%%%%%%%%%%%%%%%%%%%%%%%%%%%%%%%%%%%%%%%%%%%%%%% 
\title{Entropy production in the early-cosmology pionic phase} 
\author{Antonio Dobado}
\affiliation{Departamento de F\'{\i}sica Te\'orica I, Universidad Complutense de Madrid, 
	Parque de las Ciencias 1, 28040 Madrid, Spain.}
\author{Felipe J. Llanes-Estrada}
\affiliation{Departamento de F\'{\i}sica Te\'orica I, Universidad Complutense de Madrid, 
Parque de las Ciencias 1, 28040 Madrid, Spain.}

\author{David Rodr\'{\i}guez-Fern\'andez.}
\affiliation{Departamento de F\'{\i}sica Te\'orica I, Universidad Complutense de Madrid, 
	Parque de las Ciencias 1, 28040 Madrid, Spain.}
\affiliation{Departamento de F\'isica, Universidad de Oviedo, Avda. Calvo Sotelo 18, 33007, Oviedo, Spain.}

%%%%%%%%%%%%%%%%%%%%%%%%%%%%%%%%%%%%%%%%%%%%%%%%%%%%%%%%%%%%%%%%%% 
\begin{abstract}  
We point out that in the early universe, for temperatures in the approximate  interval 175-80 MeV (after the quark-gluon plasma), pions carried a large share of the entropy and supported the largest inhomogeneities.
Thus, we examine the production of entropy in a pion gas, particularizing to inhomogeneities of the temperature, for which we benefit from the known thermal conductivity. 
We finally put that entropy produced in relaxing such thermal inhomogeneities in the broad context of this relatively unexplored phase of early-universe cosmology.
\end{abstract} 
\date{\today}

\maketitle 

%%%%%%%%%%%%%%%%%%%%%%%%%%%%%%%%%%%%%%%%%%%%%%%%%%%%%%%%%%%%%%%%%%%%%%%%%%%%%%
\section{Introduction}
%%%%%%%%%%%%%%%%%%%%%%%%%%%%%%%%%%%%%%%%%%%%%%%%%%%%%%%%%%%%%%%%%%%%%%%%%%%%%%
The hadron and lepton phase of early-universe cosmology, spanning a temperature range between about 175 MeV and 1 MeV (between the quark-gluon plasma and cosmonucleosynthesis), has received only moderate attention in the literature~\cite{Rafelski:2013yka}, in spite of it being very rich in terms of the number of particles and interactions there present, and the underlying physics being relatively well known. This is probably because the only relic particles left from that era in the universe expansion form the cosmic neutrino background~\cite{Rafelski:2013qeu} that there is at present no hope to detect. In spite of the dearth of direct messengers from that era, it is important to pursue its study for future precision work in cosmology.

Particularly, there are few studies of the earlier part of the interval, just after exiting the quark-gluon plasma around 175-150 MeV~\cite{Brambilla:2014jmp}, when a significant fraction of the universe's entropy is carried by strongly interacting particles such as pions; only as the temperature drops below about 100 MeV, their decays $\pi_0\to \gamma\gamma$, $\pi^-\to \mu \bar{\nu}_\mu$, etc. add this entropy to that carried by leptons and photons. 

Much information on the pion phase is available from theoretical studies pertaining to the field of Relativistic Heavy Ion Collisions. Particularly, transport coefficients have been well calculated in recent years~\cite{Torres-Rincon:2012sda,Davesne:1995ms,Prakash:1993kd,Mitra:2014dia}
and can be applied to early-universe physics. This is our focuse in the present work. The particular problem that we will address is entropy production. Though most treatments assume that the universe's expansion is adiabatic and always at equilibrium, this is just the simplest hypothesis and one may fancy consider separations from that equilibrium.

One can argue that the rates of the particle-physics processes characterized in the Standard Model  are larger than the expansion Hubble factor $H=\dot{a}/a$, so the hypothesis of chemical and thermal equilibrium is reasonable, and the universe expands and cools down adiabatically. We of course concur with the analysis. But one cannot discard large past fluctuations in temperature or other quantities that have not survived to our days precisely because of the large equilibration rates damping them. So there is always a level of hypothesis involved.

What is solid information is that the fluctuations in the Cosmic Microwave Background (CMB) are measured and found small ($\nabla T/T<10^{-5}$). So one can opt for evolving large initial-state inhomogeneities so they are this small at the time of recombination, or for considering inhomogeneities that are so small in size as to evade observation in the CMB (cosmological versus microscopic inhomogeneities).
Further, since we will consider a radiation-dominated epoch, no structure-formation process is involved~\cite{Labini:2011tj}.

The largest contribution to the total entropy at zeroth order during most of the time, is due to the (quasi)massless species (photons, neutrinos and electrons), as  will be reminded below. Yet in a hot gas, since transport phenomena are diffusive, the typical transport coefficient (to which entropy production and the relaxation rate will be proportional), drops with the inverse cross-section. The case in point for this study is the thermal conductivity, $\kappa \propto 1 / \sigma$. This means that the largest inhomogeneities at a given stage of the universe evolution will be found in the gaseous subsystem which, being relativistic, is affected by the largest cross-sections. 

In the particle phase with components that are photons, leptons, and pions, the largest entropy production is thus likely to take place in the pion gas. 
Heavier hadrons are barely present already shortly after the decay of the quark-gluon plasma, for example the kaon multiplicity~\cite{Abelev:2013haa} is down by at least an order of magnitude respect to the pion multiplicity. Therefore, though in principle kaon inhomogeneities can diffuse and produce entropy, we will ignore the phenomenon altogether.

This is because their cross-sections are dictated by the strong QCD interactions and are in the 10-milibarn range, way larger than (Debye-screened, electromagnetic) lepton interactions. 
Even letting aside inhomogeneities, for $T>80$ MeV, pions actually carry a larger portion of the total homogeneous-gas entropy than photons 
(though not larger than that of leptons) because of their multiplicity, as will be shown below in figure~\ref{figure:gs} (bottom plot). 
Thus, there are two reasons to explore entropy and entropy production in the pion gas itself. In the high temperature end just after hadronization of the quark-gluon plasma, pions are  large carriers of entropy. And second, they are the ones that can support the largest inhomogeneities, if any are present, because they are the ones opposing diffusion most.

With this motivation, our concrete study will be to address the relaxation of a thermal inhomogeneity at temperature $T+\delta T$ towards the surrounding environment value of $T$, ignoring other quantities that may separate from equilibrium such as momentum distributions or chemical inhomogeneities. Then we will calculate the subsequent entropy production to have reference values that may be useful in future studies.\\

%%%%%%%%%%%%%%%%%%%%%%%%%%%%%%%%%%%%%%%%%%%%%%%%%%%%%%%%%%%%%%%%%%%%%%%%%%%%%%
\section{Entropy in the homogeneous Friedmann-Robertson-Walker cosmology}\label{sec:equilibrium}
%%%%%%%%%%%%%%%%%%%%%%%%%%%%%%%%%%%%%%%%%%%%%%%%%%%%%%%%%%%%%%%%%%%%%%%%%%%%%%
\subsection{System of equations for universe evolution}
%%%%%%%%%%%%%%%%%%%%%%%%%%%%%%%%%%%%%%%%%%%%%%%%%%%%%%%%%%%%%%%%%%%%%%%%%%%%%%
In this section we quickly review the standard statistical physics in the spatially flat ($\kappa=0$) \cite{Planck} homogeneous cosmos that serves as background for later study of inhomogeneities. In this case the two independent Einstein equations give rise to the the Friedmann equation:
\be \label{frlweq}
\left(\frac{\dot{a}}{a}\right)^2 = \frac{8\pi G}{3}\rho\ .
\ee  
for the evolution of the expansion parameter $a(t)$ from  the Friedmann equation and the balance equation
\ba \label{e2}
 \frac{d\rho}{dt}=-\frac{3\dot{a}}{a}(\rho +P)\,,
\ea
where $P$ is the total pressure and  the energy density $\rho$ is the sum of the partial energy densities for the various species
\be \label{edensity}
\rho = \rho_\gamma + \rho_{\nu,\bar{\nu}} + \rho_{e^{\pm}} +\rho_{\mu,\bar{\mu}} + \rho_{\pi^{\pm},\pi^0} + \rho_{N,\bar{N}} +\dots 
\ee
In table \ref{table:species} we summarize the main interaction channels
in the temperature range we are discussing, $1\,\text{MeV} <T<175$ MeV. In particular, for entropy considerations, nucleons (already non-relativistic) and dark matter are not important.  Pions and muons behave as radiation in the upper end of the temperature range.

This can be seen for each species $i$, with degeneracy $g_i$, contributing
\be \label{rho}
\rho_i =\frac{g_i}{(2\pi)^3}\int d^3 p\, E\,f_i({\bf r},{\bf p},t)\,,
\ee

because in thermal equilibrium, the function $f_i$ (usual Fermi-Dirac or Bose-Einstein distribution)
\be \label{distri}
f_i({\bf r},{\bf p},t)=\frac{1}{e^{(p_\alpha U^\alpha({\bf r},t) -\mu_i ({\bf r},t))/T({\bf r},t)} \pm 1}\,,
\ee
suppresses the contribution by $e^{-m_i/T}$. Here we have considered the more general case of local instantaneous thermodynamic equilibrium which will be useful later.  As usual, $p^\alpha$ and $U^\alpha$ are the components of the particle four-momentum $p^\alpha=(E,{\bf p})$ and the fluid four-velocity  $U^\alpha=\gamma_{\bf V}(1,{\bf V})$ respectively, with $\gamma_{\bf V}=(1-{\bf V}^2)^{-1/2}$. In a comoving frame, $U^\alpha=(1,0,0,0)$ and thus $p_\alpha U^\alpha =E$. We have determined the chemical potentials $\mu_i$ 
from the current abundances and the scale factor $a(t)/a({\rm today})$. As they  are tiny we do not quote them here.

\begin{table}[htbp]

	\begin{tabular}{|c @ {\hspace{5mm}}  c @ {\hspace{5mm}} c @ {\hspace{0mm}} |}
		\hline
		$\pi^0\lra \g\g$ &  $NX\lra NX$   & $e\pi\lra e\pi$ \\
		$\pi\pi\lra \pi\pi$ &  $N\bar{N}\lra \g\g\ , \pi\pi\dots$
 &$e\, e\lra e \,e$\\
		$\pi^+\lra\m^+\n_{\m}$ & $\m^+\lra e^+\n_e\overline{\n}_\m$ & $e\g\lra e\g$ \\
		$\pi\pi\lra\gamma \g$&$\m\pi\lra\m\pi$& $\g\g\lra e^-e^+$ \\
		$\n_e \overline{\n}_e\lra e^+ e^-$ & $\m^-\g\lra\m^-\g$& $\m\m\lra\m\m$ \\
		$\n_\m\overline{\n}_\m\lra \m^+\m^-$ & $\g\g\lra\m^-\m^+$ & $\m e\lra \m e$ 
%		$\overline{N}\,\overline{N}\lra \overline{N}\,\overline{N}$ 
                \\ \hline
	\end{tabular}

	\caption{Main processes in the temperature interval $175\,\text{MeV}-1\text{MeV}$. Except when specified,
the reactions can be written for all charge combinations,
 e.g. $\pi$ denoting either of $\pi^+$, $\pi^-$ or $\pi^0$, and $\m$ both muon and antimuon. 
(We omit additional reactions with much smaller branching fractions, such as  the Dalitz decay $\pi^0\to\g\g^*\to e^+ e^-\g$ at $O(1\%)$ \cite{part}.)
In this article we focus on the temperature range where $\pi\pi\lra\pi\pi$ dominates transport.
} \label{table:species}
\end{table}

Summing Eq.~(\ref{rho}) over species yields $\rho(T)$ from which the temperature evolution 
can be extracted as
\be \label{eqforT}
\frac{dT}{dt}= -\frac{3\dot{a}}{a}(\rho +P)\frac{dT}{d\rho}\, .
\ee
 Eq.~(\ref{frlweq}) and~(\ref{eqforT}) can be solved numerically by using for example the Runge-Kutta algorithm.
The energy density is 
computed from the numerical integration of  Eq.~(\ref{rho}) at each temperature, and the pressure is similarly obtained from the spatial trace of the energy-momentum tensor $\delta_{jk} T^{jk}=3P$, 
which results in a  sum over partial pressures of all species,
\be \label{pressure}
P = \sum_i P_i =\sum_i \frac{1}{3}\int d^3 p\, f_i({\bf r},{\bf p},t) \frac{\vert {\bf p}\vert^2}{E}\, .
\ee

Density and pressure decrease monotonically with $t$, while the scale factor $a(t)$ increases monotonically; any of them may be used as a clock for further computations. We will set as origin of time the exit from the QGP at the top of the temperature interval, 
$0\equiv t_{T=175{\rm MeV}}$,
where we set $a(0)\equiv 1$.

With the solutions at hand we can backtrack from the time of nucleosynthesis (a well-studied period~\cite{Burles:2000zk}) to the pion gas at temperatures two orders of magnitude higher, since the entire particle content in this epoch is well known. We do not resort to usual textbook power-law approximations since simple computer codes produce the 
(numerically) exact solutions for this one-dimensional evolution-equation set. For computer accuracy, it is necessary to set a unit system that minimizes the number of large powers. We often take $(100\,\text{MeV})$ for temperature, energy and chemical potential ($k_B=1$) and $\text{peV}^{-1}$ for time and space ($c=1$). With this, the Cavendish constant $G$ turns out to be $1/1.44\,(100\,\text{MeV})^2$. Dimensionally,  time is an inverse energy, so that 
$1\,\text{s} \rightarrow 1.52\times 10^{3}\, \text{peV}^{-1}$.

%\frac{1\,\text{s}}{6.58\times 10^{-16}\text{eV}\cdot s}= 1.52\times 10^{15} \,\text{eV}^{-1}= 
% likewise, $10^{-6}\,\text{s}$ can be recast as $1.52\times 10^{-3}\,\text{peV}^{-1}$.\\

\begin{figure}
\includegraphics[scale=0.65]{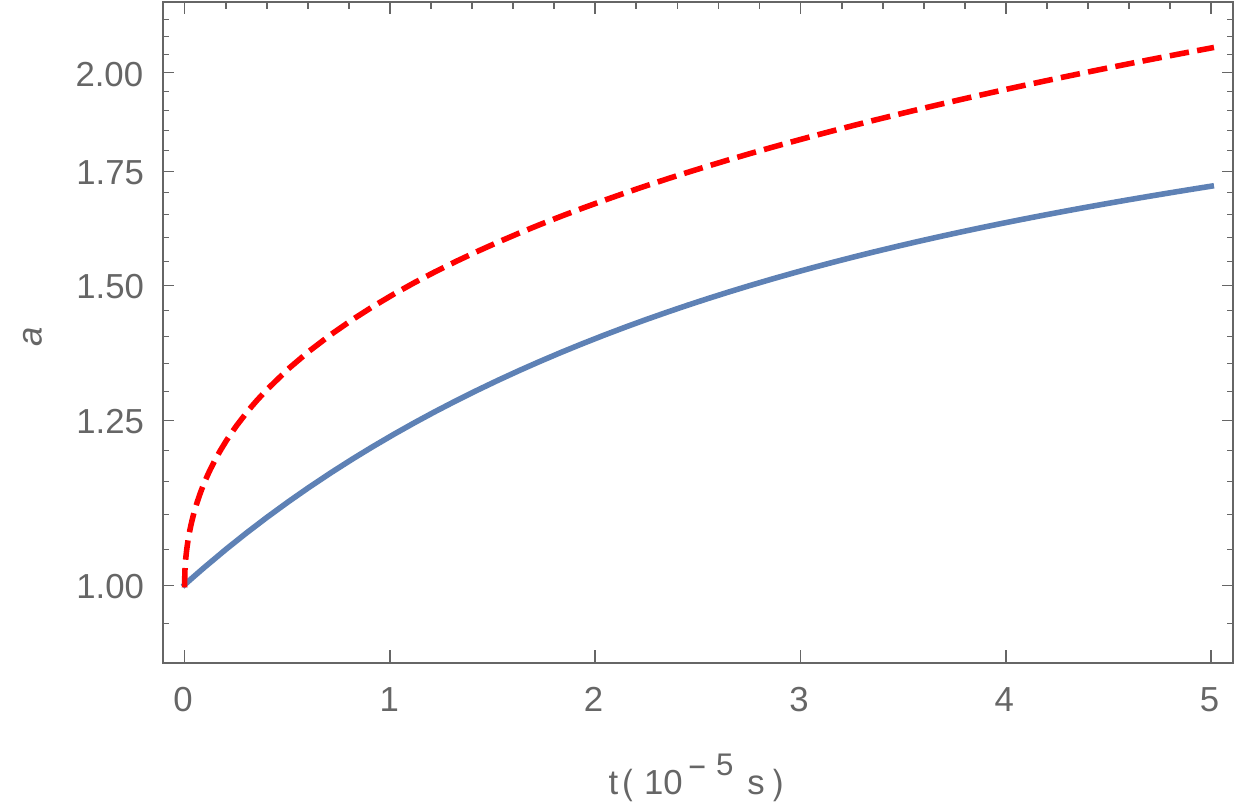}
\caption{Computed scale factor as a function of time (solid, blue online). We take as normalization, $a(0)=1$, where $T(t=0)=175$ MeV.
We also show the analytic form $a(t) \propto \sqrt{t}$ (dashed line, red online) that fits the lepton era at smaller temperatures. As we focuse here on the pion gas, found at earlier times, roughly from $t_{T=175}=0\,\text{s}$ to $t_{T=100}\approx 10^{-2}\, \text{peV}^{-1}\, (10^{-5}\,\text{s})$, 
that square-root approximation separates significantly from the actual numerical computation that we employ.
}
\label{figure:scale}
\end{figure}

The resulting scale factor is shown in figure~\ref{figure:scale}.\\

\vspace{-6mm}
%%%%%%%%%%%%%%%%%%%%%%%%%%%%%%%%%%%%%%%%%%%%%%%%%%%%%%%%%%%%%%%%%%%%%%%%%%%%%%%%%%%%%%%%%%%%%%%%%%%%%
\subsection{Computation of the entropy}
%%%%%%%%%%%%%%%%%%%%%%%%%%%%%%%%%%%%%%%%%%%%%%%%%%%%%%%%%%%%%%%%%%%%%%%%%%%%%%%%%%%%%%%%%%%%%%%%%%%%%

For the calculation of the entropy in the homogeneous case we first note that the thermodynamic magnitudes are a function of the temperature only. We will also assume thermal equilibrium, vanishing chemical potentials and adiabatic expansion. Then the conservation of the entropy per co-moving volume implies $s a^3=s_0 a_0^3$ where $s=s(T)$ is the entropy density (see for example \cite{Weinberg}).

\begin{figure}
\includegraphics[scale=0.55]{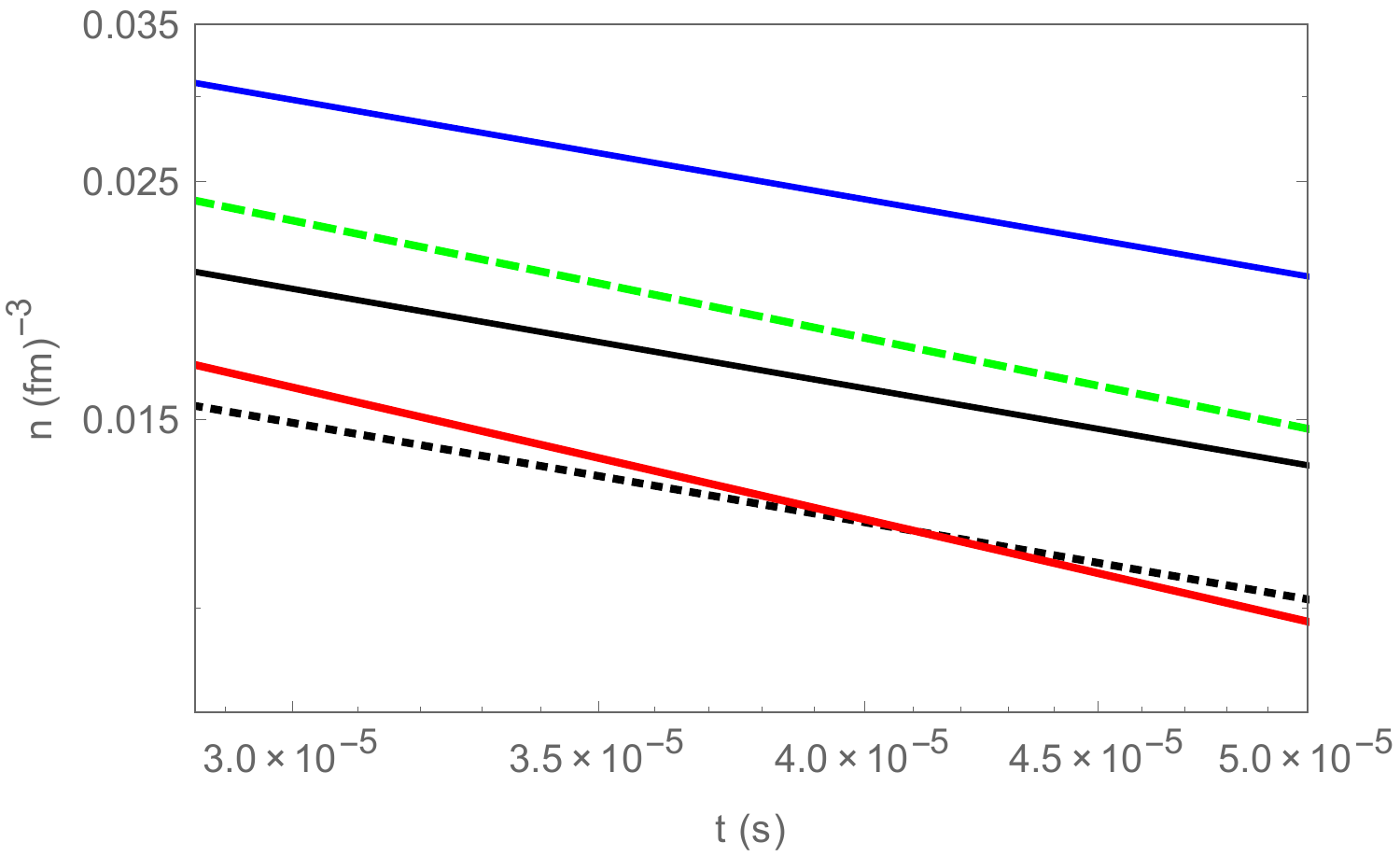}
\caption{ \label{figure:densityplot}
(Color online).
Number densities of the most abundant species during the hadron-lepton epoch. 
From top to bottom:
$e^-, \ e^+$ (blue); $\mu^-,\ \mu^+$ dashed line (green); $\gamma$ (black); $\pi$ (red, solid); 
$\nu$ (black, dotted).
The number density of nucleons is completely negligible during the entire time interval.
}%end caption
\end{figure}

The second principle of thermodynamics gives the total-entropy increase as
\be \label{termo2}
TdS = d(\rho V)+P dV .
\ee
where $S=s V$ is the total entropy and $V$ is the volume. From this equation it is possible to get the thermodynamic relations:
\be
s  =  \frac{1}{T}(\rho+P)= \frac{dP}{dT}.
\ee
Therefore the entropy in  co-moving volume $V$ is constant and proportional to $a^3(t)\frac{\rho +P}{T}$.  
Enumerating all the species (in equilibrium at the same universal $T$)
\be \label{entropy0}
s=\frac{\rho_1+\cdots +\rho_n +P_1+\cdots +P_n}{T}.
\ee
The pion gas can be near chemical equilibrium because the pion production rate (through $\gamma\gamma \leftrightarrow \pi \pi$ followed by $\pi^+\pi^-\leftrightarrow \pi^0\pi^0$, and similar lepton-lepton inelastic interactions) is sufficient to offset the pion decay rate.
In figure~\ref{figure:densityplot} we show the time-evolution of the number density for the more relevant temperature span between 175 and 70 MeV. During this time interval, pions (and also muons) are abundant, comparably to the (quasi)massless species.

The entropy density may also be written as $\propto g_s(T)\,T^3$, $g_s(T)$ being the number of effective degrees of freedom. In particular, for ultrarelativistic particles,
\be 
s(T)=g_s\frac{2\pi^2}{45}T^3\,.
\ee

\begin{figure}
\includegraphics[scale=0.65]{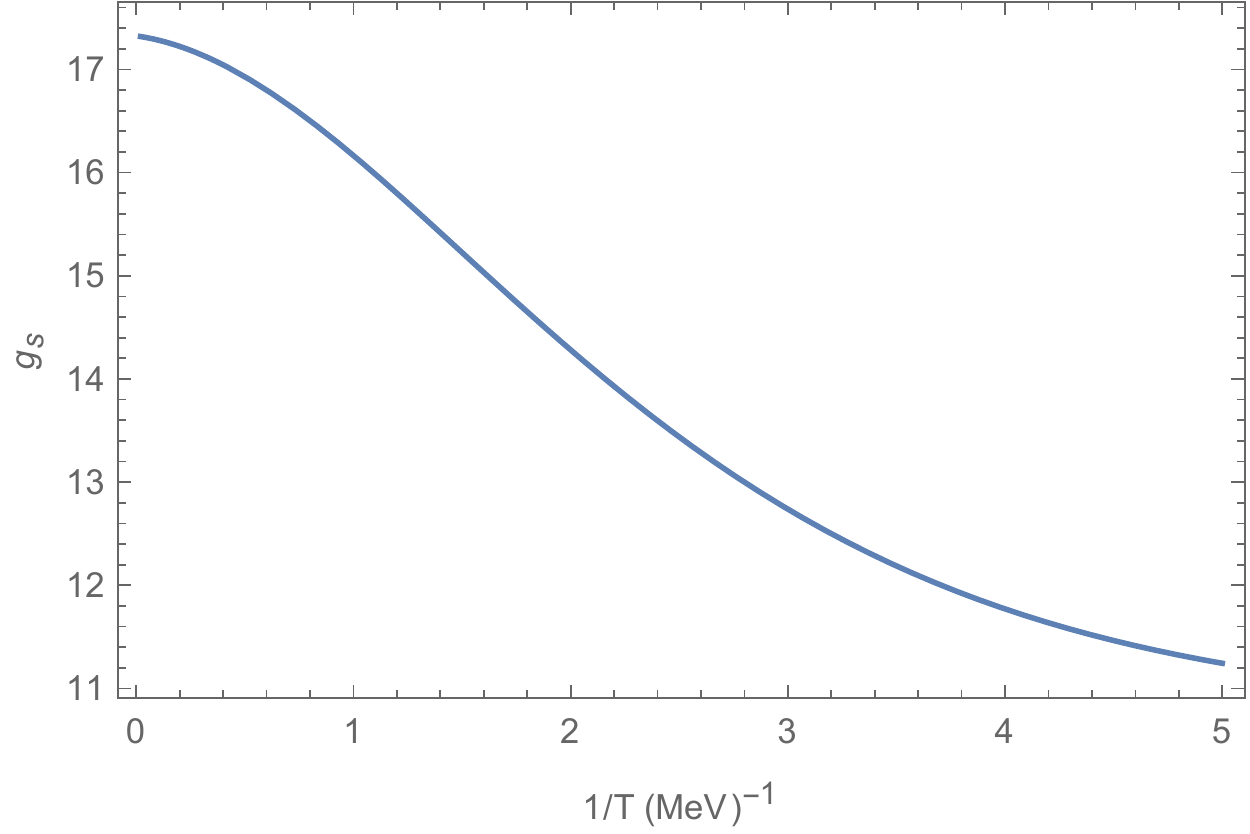}
\vspace{3mm}
\includegraphics[scale=0.65]{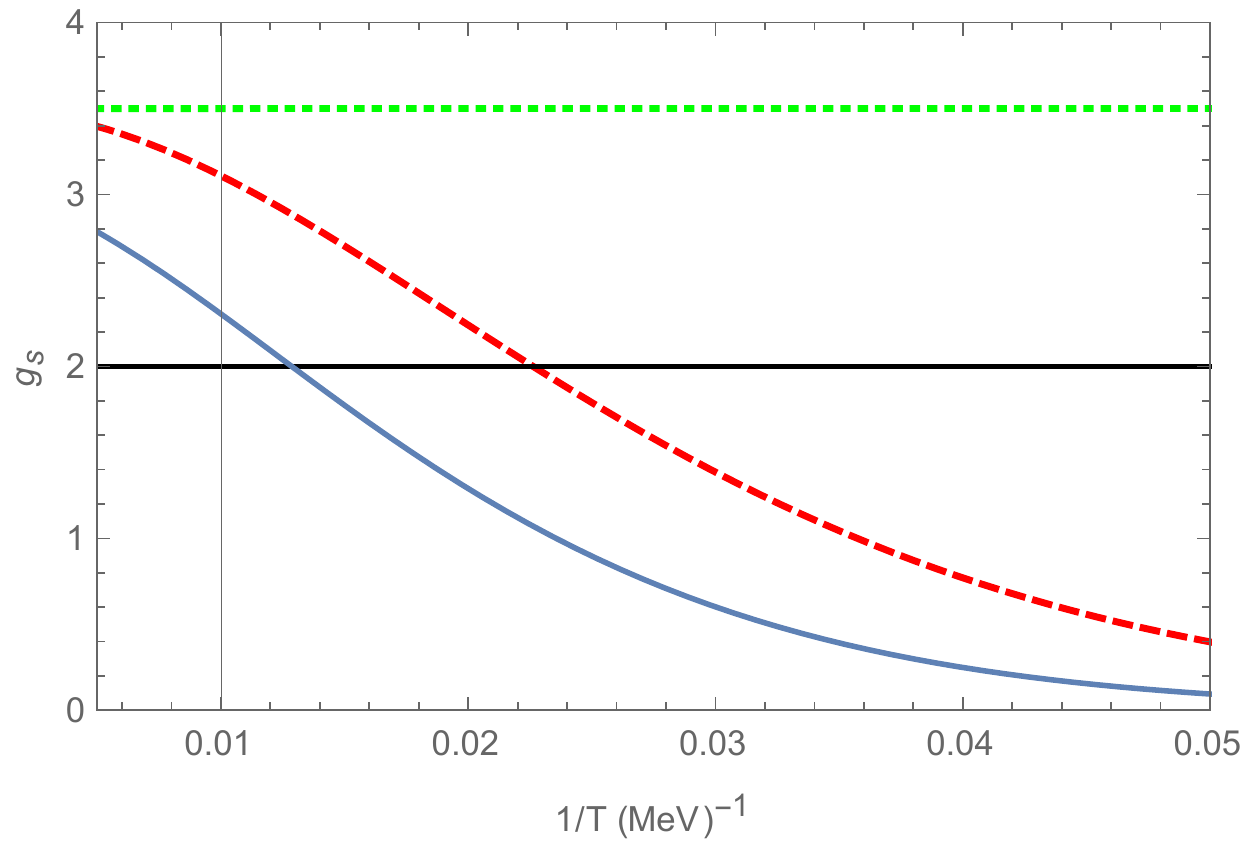}
\caption{Top plot: aggregated effective number of relativistic degrees of freedom $g_s$ as a function of $\beta = 1/T$ from numerical calculation. Bottom plot: effective number of degrees of freedom for pions (solid line, blue online), photons (horizontal solid line, black online), electrons and positrons (dotted horizontal line, green online) and muons (dashed line, red online). Note that at the highest part of the temperature interval, pions provide a larger contribution to the entropy density than photons, though leptons are the largest carriers of entropy. 
\label{figure:gs}}
\end{figure}

Due to their relativistic behavior
throughout our entire temperature range, the effective number of degrees of freedom for photons, electrons and neutrinos is constant. The massive species see drops when $T<m_i$ as they  become non-relativistic. Our numerical computation of the entropy can be casted in terms of $g_s(T)$ and it is plotted in figure~\ref{figure:gs}. In particular, the contribution of nucleons (as well as all strange and higher-flavor particles, not mentioned further) to the entropy density is completely negligible.

\vspace{-6mm}

%%%%%%%%%%%%%%%%%%%%%%%%%%%%%%%%%%%%%%%%%%%%%%%%%%%%%%%%%%%%%%%%%%%%%%%%%%%%%%
\section{Entropy production by local departures from (thermal) homogeneity}
%%%%%%%%%%%%%%%%%%%%%%%%%%%%%%%%%%%%%%%%%%%%%%%%%%%%%%%%%%%%%%%%%%%%%%%%%%%%%%
\subsection{Solution to the heat equation}
%%%%%%%%%%%%%%%%%%%%%%%%%%%%%%%%%%%%%%%%%%%%%%%%%%%%%%%%%%%%%%%%%%%%%%%%%%%%%%
In this section, we consider separations from the homogeneous background described in section~\ref{sec:equilibrium}.
For simplicity we will take inhomogeneities to be spherical bulbs at  temperature different from the background.
Thus, the temperature field $T({\bf r},t)$ will now also acquire a dependence on position. Local thermal equilibrium as well as chemical equilibrium is still assumed (and departures thereof can be separately considered in further investigations that we do not attempt here).

\begin{figure} 
\includegraphics[scale=0.65]{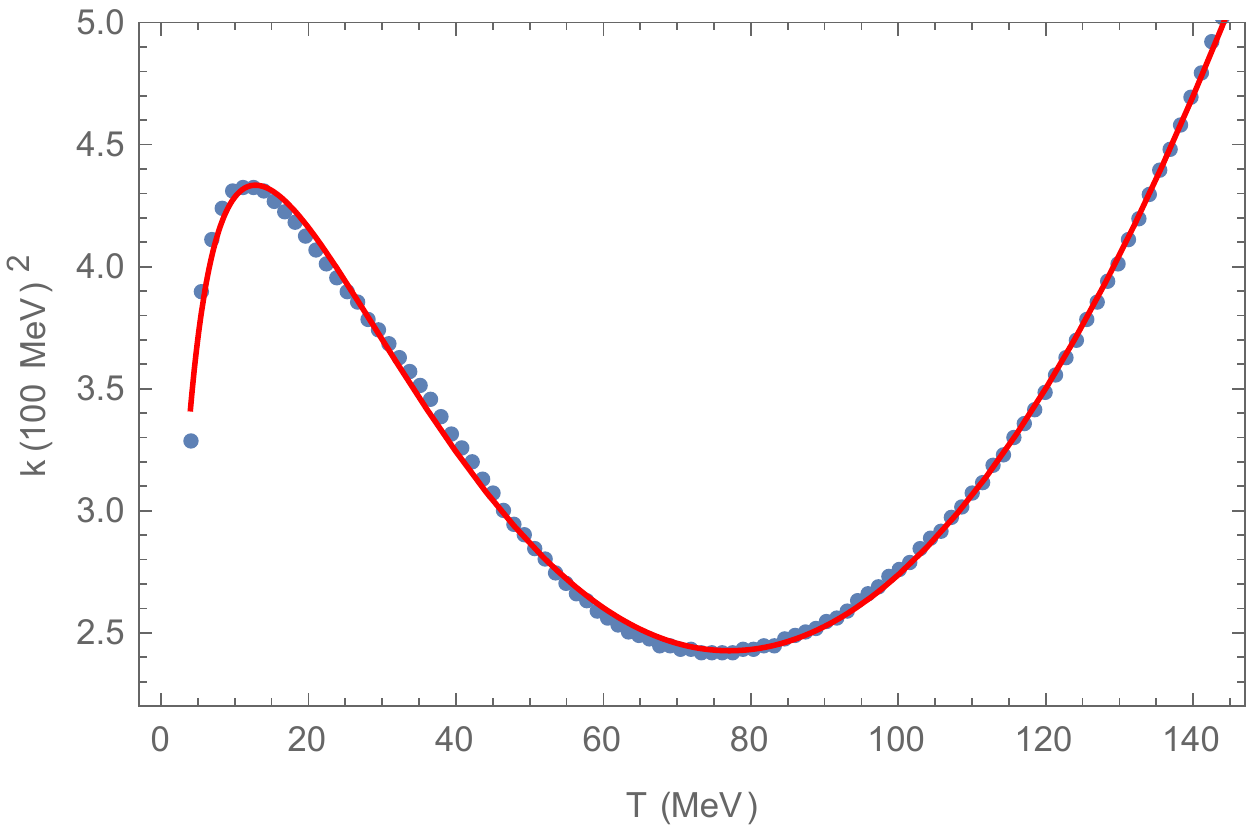}
\caption{Dotted blue online: Thermal conductivity $\kappa$ as function of temperature at zero chemical potential from solving the Boltzmann equation. Solid (red online):  simple interpolating function 
employed in the heat equation solver. 
\label{figure:k}}
\end{figure}

The departure from the background modifies temperature and entropy density
\ba \label{inhom}
 T({\bf r},t) &=&  T_{back}(t) +\delta T ({\bf r},t)\,,  \nonumber  \\
 s({\bf r},t) &=&  s_{back}(t) +\delta s({\bf r},t)\, .
\ea
Setting as simplest initial condition a bubble of higher $T$ than the surroundings, the temperature profile of such bulb will evolve according to the heat equation. 
Then,
\be \label{heatin}
\Delta \left( \delta T({\bf r},t)\right) = \frac{\kappa(T)}{c_p(T)}\, 
\frac{\partial \left(\delta T(\bf{r,t})\right)}{\partial t}\,.
\ee
with $\kappa(T)$ being the heat conductivity. Here the constant-pressure specific heat $c_p$ is defined as the derivative of the background entropy (neglecting the newly produced one) 
with respect to temperature at constant $P$: 
\be \label{cp}
c_p(T) =\frac{\partial s_{back}(T)}{\partial T}\Bigg\vert_P\, .
\ee
Since we already calculated the contribution of pions to the entropy density $s^\pi_{back}$ we can immediately compute the partial specific heat of the pion gas (we will further drop the superindex $\pi$ in this section, as all quantities are refered to the pion gas alone).  
The other non-trivial function is $\kappa(T)$, the thermal conductivity, which depends on the temperature alone and is known from recent and earlier studies. The numeric data~\cite{Torres-Rincon:2012sda}
 from a variational solution of Boltzmann's equation following the Chapman-Enskog expansion is shown in figure~\ref{figure:k}.
Since $c_p(T)$ and $\kappa(T)$ are nontrivial functions of the temperature, the heat equation does not admit an immediate analytical solution, so we numerically solve it by brute force with the simplest parabolic solver for a partial differential equation based on the finite-difference method in space and the Euler method in time. Thus, in figure~\ref{figure:k} we also show a simple interpolating function for the conductivity in the temperature interval of interest that we employ to speed up the computer code.

The valley in the conductivity at mid-temperatures occurs because of the $m_\pi\simeq f_\pi$ scales; the dropping low-temperature behavior can be obtained from the $\pi\pi$ scattering length and non-relativistic kinetic theory, and at high-$T$ dimensional analysis dictates $\kappa\propto T^2$ as visible. The detailed calculation with the full machinery of phase shifts, unitarity, chiral perturbation theory, etc. has been reported elsewhere \cite{Torres-Rincon:2012sda}.

\begin{center}
\begin{figure}
\includegraphics[scale=0.65]{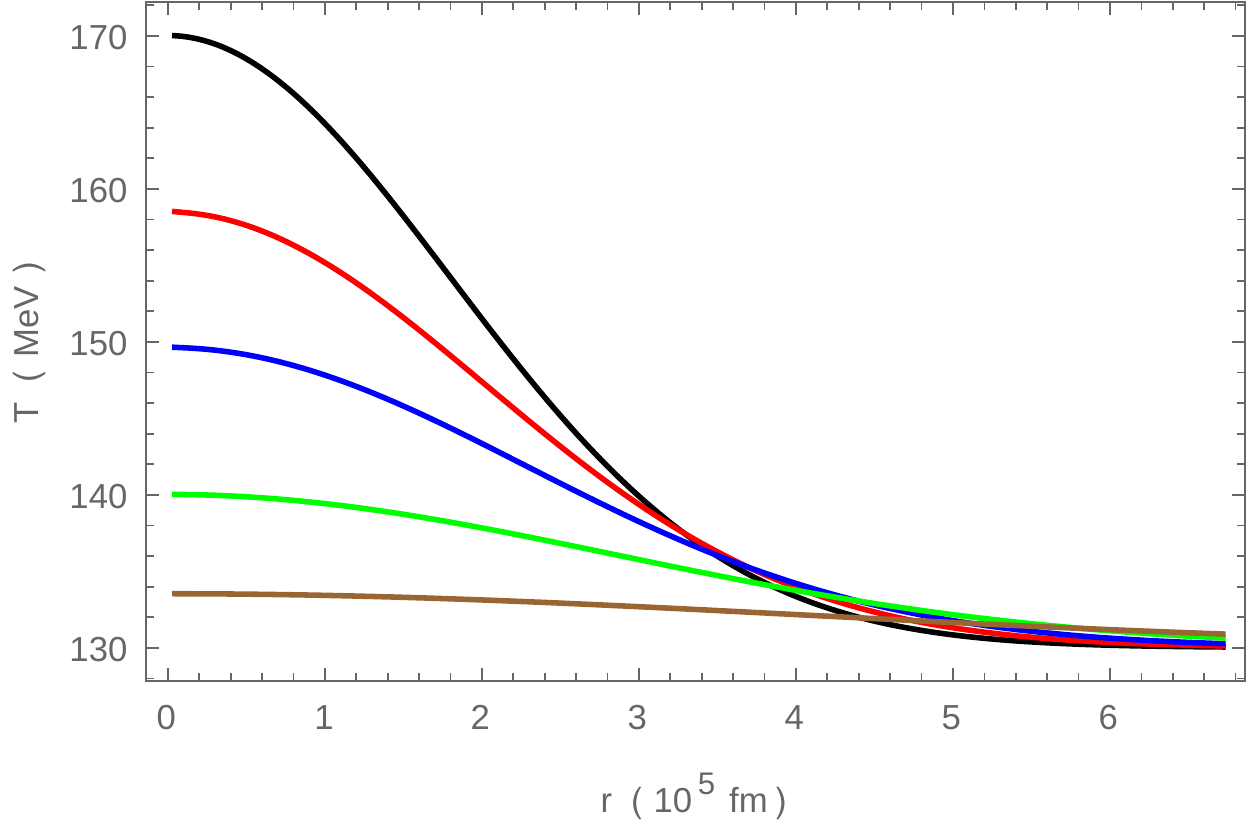}
\caption{Temperature profile $T(r,t)$ of an inhomogeneity of initial size $R=2.5\times 10^5\,\text{fm}$, as a function of the radius $r$ for increasing times. Top, solid black line: initial condition $T(0,0)=170 \,\text{MeV}$. Brown solid line, much flatter of the bottom: $T(r,t_r)$, $t_r\approx 10^{-12}\,\text{s}$. Other lines illustrate the time evolution of the inhomogeneity at intermediate times.
\label{figure:deltat}}
\end{figure}
\end{center}

The numeric solution of  Eq.~(\ref{heatin}), $\delta T({\bf r},t)$, is shown in figure \ref{figure:deltat} with an initial condition that has a spherical profile Gaussian in the radius,
\be \label{init}
 \delta T(r,0)=\delta T_0\, e^{- \frac{r^2}{2R^2}}\,.
\ee
Here $\delta T_0$ is the initial central temperature of the inhomogeneity over that of the background, and $R$ is the typical radius.

There are several considerations to choose the size of the inhomogeneity. At the largest scale,
we can ask ourselves what is the largest possible radius that will homogenize during the pion gas lifetime. We must also take the size of the bulb small enough so as to respect CMB constraints.
%As can be glanced back in figure~\ref{figure:scale}, 
%the scale factor can be nicely fitted by $a(t)\propto \sqrt{t}$. Thus, 
Glancing back to figure~\ref{figure:scale}, we estimate 
the Hubble horizon reached during the pion gas to be around $10^{-3}-10^{-2}$ peV$^{-1}$. This means that no homogenization can take place over distances larger than about a light second $(1-10)\times 10^{-3}$ peV$^{-1}$, or squaring and inverting,  $R$ must be no larger than $\approx 10^{16}-10^{17} \text{fm}$. This guarantees that the thermal flattening of the bulb never violates causality.
Further, since the first order heat equation is not relativistically causal and we have not examined the $2^{nd}$ order formalism, we have to restrict ourselves to even significantly smaller spheres.
A further consideration is that if the inhomogeneity is too large, its relaxation time will be so great that when it reaches thermal equilibrium, there are no pions left (they are abundant for $T_{back}\approx 175-80$ MeV). For this reason (exclusively of simplicity), we will restrict the study to inhomogeneities no bigger than $R \approx 10^9$ fm. 
These are small enough not to perturb the metric significantly, so we can treat them simply as Newtonian perturbations.
Finally, when we consider the smallest radii of the inhomogeneity, in the typical nuclear scale or somewhat more, RHIC guidance is available.

%%%%%%%%%%%%%%%%%%%%%%%%%%%%%%%%%%%%%%%%%%%%%%%%%%%%%%%%%%%%%%%%%%%%%%%%%%%%%%
\subsection{Entropy increase in one inhomogeneity}
%%%%%%%%%%%%%%%%%%%%%%%%%%%%%%%%%%%%%%%%%%%%%%%%%%%%%%%%%%%%%%%%%%%%%%%%%%%%%%

The variation of the entropy of our inhomogeneity of volume $V$ during the relaxation process can be written as: 
\be \label{inentropy}
 \frac{dS_T}{dt}=\frac{dS_{\bar{V}}}{dt} +\frac{dS_V}{dt}\,,
\ee
where $S_T$ denotes the total entropy, $dS_{\bar{V}}$ represents the  entropy exchanged with the rest of the universe and $dS_V$ the inner entropy production.   The exchanged entropy $dS_{\bar{V}}$ can be obtained by means of an integral of the incomming entropy current over the surface of the inhomogeneity $\partial V$. We will consider the exchange as positive if entropy is supplied to the subsystem by the surroundings. The entropy current will be denoted by ${\bf j}_s$. 
Concerning the internal entropy production $d S_V$ we introduce  the   rate of entropy production $\sigma_s$  per unit volume and unit time inside the system. In terms of these quantities, $dS_{\bar{V}}/dt$ and $dS_V/dt$ may be written as 
\ba \label{entropye}
\frac{dS_{\bar{V}}}{dt} &=& -\int_{\partial V}  {\bf j}_s\cdot {\bf n} \,d\Sigma\,,  \nonumber\\
\frac{dS_V}{dt} &=& \int_V \sigma_s \, dV\, .
\ea

 Expressing Eq.~(\ref{inentropy}) in terms of the entropy current and density we have: 
\be \label{enprod}
\frac{d S_T}{dt} = \frac{d}{dt}\int_V   s_T\,  dV =-\int_{\partial V} d\Sigma\,\,{\bf j}_s\cdot {\bf n} +\int_V \,dV\,\sigma_s\ 
\ee
and use of Gauss's theorem yields the  equation
\be
\label{enprod2}
\frac{d s_T}{dt} = -\boldsymbol{ \nabla}\cdot {\bf j}_s +\sigma_s\ .\\
\nonumber
\ee 
 For small flows, linear laws hold, such as the Fourier law for the heat flux: 
\be\label{Fourier}
{\bf j}_e=-\kappa(T)\boldsymbol{ \nabla} T\ ;
\ee 
where  ${\bf j}_e$ is the heat current vector. Other examples of linear laws are Fick's law for a flavor $i$ concentration flux, ${\bf j}_i=-D_i\boldsymbol{ \nabla} n_i$, with $D_i$ being a  diffusion coefficient for the particle species
$i$; or Ohm's law for the electric current density ${\bf j}_Q =- \kappa \boldsymbol{ \nabla}\phi$ with ${\bf j}_Q$ being the electric current, $\phi$ the electric potential and $\kappa$ the electric conductivity.
A general form for the entropy production $\sigma_s$ is
\be 
\sigma_s= {\bf j}_e\cdot \boldsymbol{ \nabla}\left(\frac{1}{T}\right)-\sum_i \left[ {\bf j}_i\cdot \boldsymbol{ \nabla}\left(\frac{\mu_i}{T}\right) +\frac{A_k\,v_k}{T}\right]+\kappa \frac{{\bf I}\cdot {\bf j}_Q}{T}\cdots \,,
\ee
with $A_k,v_k$ the activities and the  stoichiometric coefficients  for the $i$th species involved in inelastic particle reactions.  In the following we will consider  the entropy production $\sigma_s$ for the thermal flow alone (first term). 
Basic thermodynamics yields
\be 
dU = T\, d S_T =\left({\bf j}_e\cdot {\bf n}\right) d\Sigma\, dt
\ee
where $U$ is the internal energy of  the inhomogeneity. Integrating over the surface and in time, and using Gauss's theorem, we find the entropy produced in the process of relaxation of the inhomogeneity: 
\be 
\Delta S_T =\int _{\partial V} \,d\Sigma\, dt \,\frac{ {\bf j}_e\cdot {\bf n}}{T}=\int _V \,dV\, dt   \boldsymbol{ \nabla} \cdot  \left(    \frac{ {\bf j}_e}{T}\right)
\ee
Applying Fourier's law in Eq.~(\ref{Fourier}) we find:
\be 
\Delta S_T = \int \,dV\,dt \boldsymbol{ \nabla}\cdot \left(-\frac{1}{T}\kappa(T) \boldsymbol{ \nabla} T\right) \ .
\ee

Applying now Leibnitz's rule we get 
\ba \label{eprod}
\Delta S_T = \int dV\, dt\,\frac{\kappa(T)}{T^2}
\left(\vert \boldsymbol{\nabla}\delta T\vert ^2 -T    \Delta \delta T\right)\, ,\nonumber\\
\ea
which is positive, $\Delta S_T \geq 0$, since $ \boldsymbol{\Delta} T = \boldsymbol{ \Delta}(\delta T)\leq 0$ (remember that $T_{back}$ was position-independent).
  
Comparing with Eq.~(\ref{entropye}) we find the production of entropy and the divergence of its flow

\ba \label{ents}
\sigma_s(r,t) &=& \frac{\kappa(T)}{T^2}\vert \boldsymbol{\nabla} \delta T(r,t)\vert^2\,,\\
-\boldsymbol{ \nabla}\cdot {\bf j}_s(r,t) &=& \frac{\kappa(T)}{T} \Delta \delta T(r,t)\,.\\
\nonumber
\ea

The internal entropy produced in dissipating an inhomogeneity is an integrated entropy $\Delta S_V$, obtained from the entropy-density production $\sigma_s$ after integrating over the time and space when and where the inhomogeneity was relevant, 
\be \label{intentropy}
\Delta S_V(\delta T_0)=\int \,dV\,dt \, \sigma_s(r,t)\,.
\ee
To ascertain the size of this produced entropy and assess its relative importance, it is natural to quotient it by the background entropy in the same volume, $S_{back}$, that for a spherical disturbance integrating up to the radius $R$ (defined above in Eq.~(\ref{init}) as the characteristic Gaussian fall-off radius) is

\be \label{sback}
S_{back}(R,T_{back})
% =V_s \cdot s_\pi 
\simeq \frac{4}{3}\pi R^3 s_{back}(T_{back})\,.
\ee

We now have all necessary equations and can proceed to the numerical computation.

%%%%%%%%%%%%%%%%%%%%%%%%%%%%%%%%%%%%%%%%%%%%%%%%%%%%%%%%%%%%%%%%%%%%%%%%%%%%%%
\section{Numerical results}
%%%%%%%%%%%%%%%%%%%%%%%%%%%%%%%%%%%%%%%%%%%%%%%%%%%%%%%%%%%%%%%%%%%%%%%%%%%%%%
\subsection{One inhomogeneity only}
%%%%%%%%%%%%%%%%%%%%%%%%%%%%%%%%%%%%%%%%%%%%%%%%%%%%%%%%%%%%%%%%%%%%%%%%%%%%%%

To check the computer codes and understand the typical order of magnitude, let us consider a time period that is short enough so that the background temperature does not vary appreciably and can be considered constant ($T=T_0$). That means in particular that  $\kappa$ and $c_p$ also remain constant  (in fact the inhomogeneity has not fully spread in this case, but we can deal with this numerically later). Then we can make the replacement 
\be 
\sigma_s  \simeq \frac{\kappa( T_0)}{ T_0^2\,R^4}\delta T_0^2r^2 e^{-\frac{r^2}{2R^2}}\,,
\ee
wherein $T_0 =T_{back}+\delta T_0$. For a temperature interval from $175$ MeV to $170$ MeV ($t\in [0-10^{13}]$  MeV$^{-1}$), we keep $\kappa(T)/T_0^2$ unchanged and of order one, thus, $\sigma_s\propto \frac{\delta T_0^2}{R^4}r^2 e^{-\frac{r^2}{2R^2}}$. Carrying out the integral over space, one gets
\be 
\int\,dV\sigma_s\propto \delta T_0^2\,R\,.
\ee

To put some numbers, take an inhomogeneity of size 
$ 10^{8}$ fm at $\delta T_0 \approx 10 \,\text{MeV}$; one has then $\int\,dV \sigma_s\simeq 10^{7}$  MeV. 
This element multiplied by a time interval $\Delta t\approx 10^{13}$ MeV$^{-1}$ gives an integrated entropy of order $10^{20}$. Nevertheless, since at $175$ MeV $s_{back}$ is numerically of order $10^6$ MeV, the background entropy $S_{back}$ given by Eq.~(\ref{sback}) is $\approx 10^{24}$, so the ratio $\Delta S_V/S_{back}$ is $\approx 10^{-4}$. Inasmuch as we are considering just a tiny time interval in which the bubble did not have enough time to evolve,  the value of the entropy produced over the entire life of the bubble must be larger than this figure, and thus not negligible at all (but requires a numeric computation).

Now, by solving  the heat equation for $T(r,t)$ in Eq.~(\ref{heatin}), we can compute the integral in Eq.~(\ref{intentropy}) with Eq.~(\ref{ents}) and thus numerically obtain  
$\Delta S_V$.

\begin{table}[htbp]
\caption{Values of $\Delta S_V/S_{back}$ (in units of $10^{-3}$) for different temperatures and, for each column, a  value of the inhomogeneity size given by $R_1=2.5\times 10^{9}\,,R_2=2.5\times 10^{7}\,,R_3=2.5\times 10^5\,,R_4=2.5\times 10^2$	in fm units. Temperatures are in MeV. \label{table:entropy}}
\begin{tabular}{|c @ {\hspace{2mm}} c @ {\hspace{5mm}} | c @ {\hspace{5mm}} c @ {\hspace{5mm}} c @ {\hspace{5mm}} c @ {\hspace{0mm}}| }
\toprule
$T_{back}$ & $\delta T_0$ &  $1$ &$2$ & $3$ &$4$\\
\toprule
\multirow{8}{0.35cm}{130} 

& 40 & 56.6 & 46.2 & 46.2 & 43.6 \\
& 35 & 46.7 & 35.6 & 35.6 & 33.6\\
& 30 & 37.7 & 26.3 & 26.3 & 24.8\\
& 25 & 30.1 & 18.4 & 18.4 & 17.4\\
& 20 & 24.6 & 11.8 & 11.8 & 11.1\\
& 15 & 19.8 & 6.7 & 6.7 & 6.3 \\
& 10 & 15.9 & 3.0 & 3.0 & 2.8\\
& 5  & 18.9 & 0.8 & 0.8 & 0.7\\
  \hline
  
\multirow{8}{0.35cm}{100} 
& 40 & 162.3 & 132.7 & 132.7 & 125.0 \\
& 35 & 132.6 & 101.9 & 101.1 & 95.3 \\
& 30 & 105.6 & 73.9 & 73.9 & 70.0\\
& 25 & 83.6 & 51.0 & 51.0 & 48.1\\
& 20 & 67.5 & 32.4 & 32.4 & 30.6\\
& 15 & 53.5 & 18.0 & 18.0 & 17.1\\
& 10 & 42.5 & 7.9 & 7.9 & 7.5\\
& 5 & 48.7 & 2.0 & 2.0 & 1.9\\

\toprule
\end{tabular}
\end{table}

\begin{figure}
\includegraphics[scale=0.65]{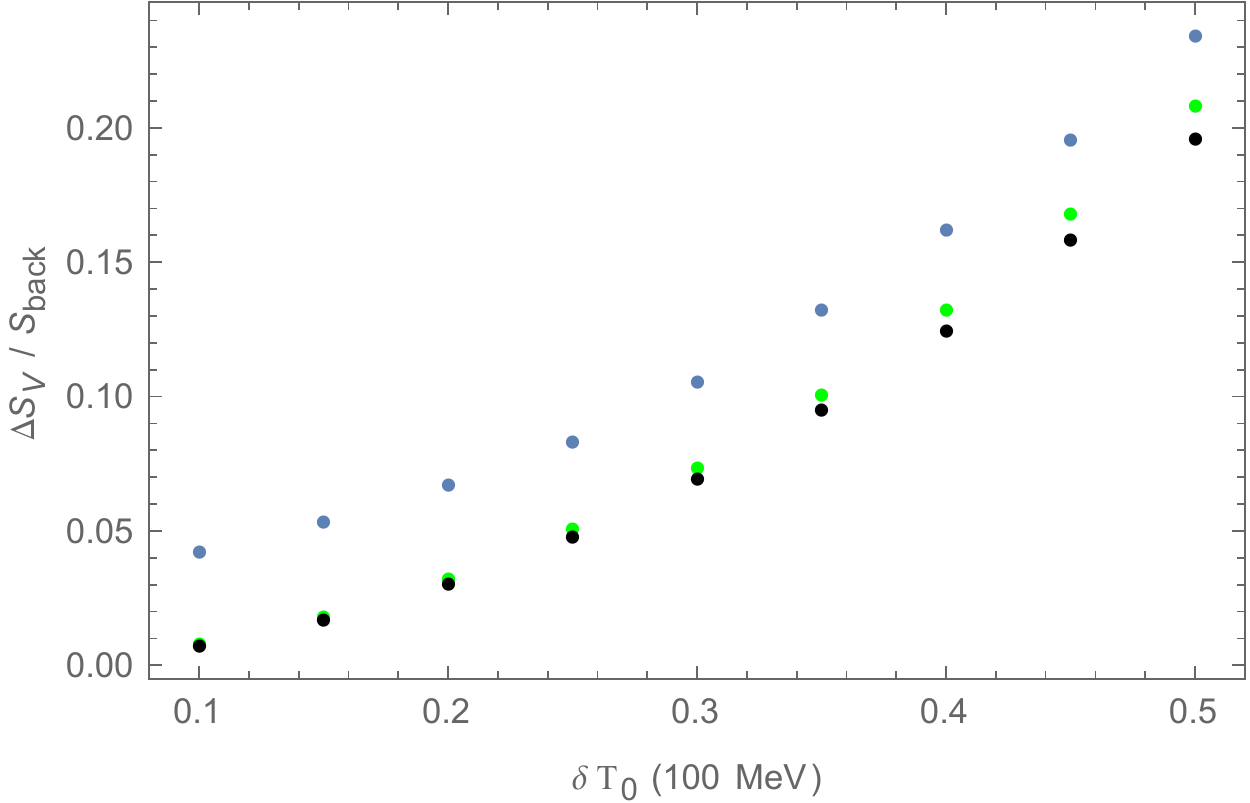}
\caption{$\Delta S_V/S_{back}$ against $\delta T_0$ for $T_{back}=100$ MeV and different radii $R$, from top to bottom: $R=2.5\times 10^9$ fm (blue online), $R=2.5\times 10^5$ fm (green online), and $2.5\times 10^2$ fm (black online). \label{fig:entdt} }
\end{figure}

\begin{center}
\begin{figure}
\includegraphics[scale=0.28]{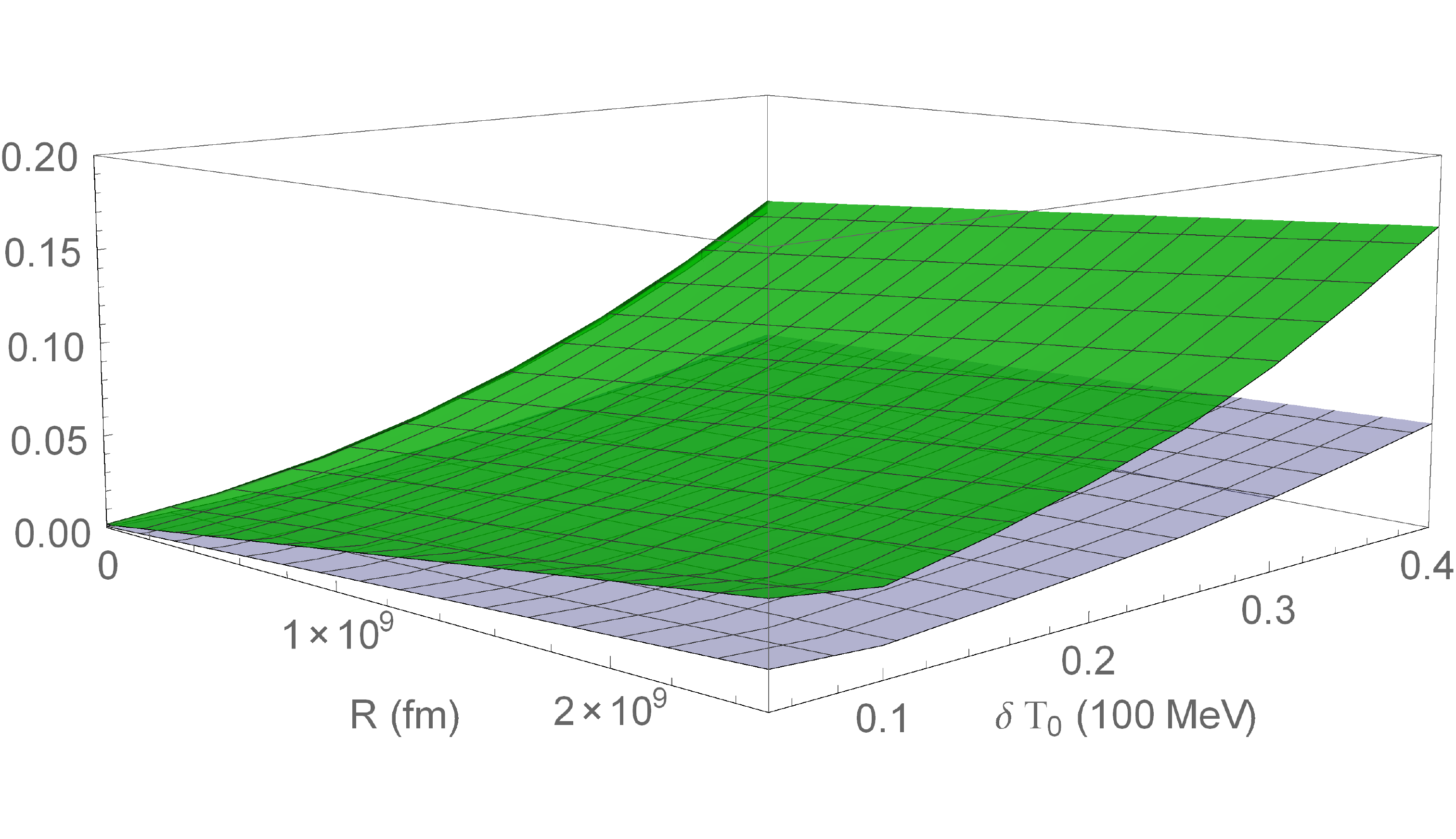}
\caption{$\Delta S_V/S_{back}$ as a function of $R$ and $\delta T_0$ at different $T_{back}$ in the moment of the formation of the inhomogeneity. From top to bottom the surfaces correspond to $T_{back}=100$ MeV (green online) and $T_{back}=130$ MeV (blue online). \label{fig:entone}}
\end{figure}
\end{center}

Table \ref{table:entropy} shows the numeric computation of $\Delta S_V$ divided by $S_{back}$ for different choices of $T_{back}$ and $\delta T_0$ (initial, central intensity of the perturbation).
 In figure \ref{fig:entdt} we plot the same quantity $\Delta S_V/S_{back}$ against $ \delta T_0$ for different initial sizes. As expected, the bigger the bulb is, the more entropy it produces, also relative to the background.
In figure \ref{fig:entone} we  simultaneously plot $\Delta S_V/S_{back}$ against the size $R$ and intensity $\delta T_0$ of the inhomogeneity. It is interesting though expected to note that at lower background temperatures the integrated entropy becomes larger for equal $\delta T_0$. 
Mathematically this comes from the term $\vert\boldsymbol{\nabla}T\vert^2$ in Eq.~(\ref{ents}), 
which increases as the $T_{back}$ decreases, giving rise to a larger entropy production.\\

%%%%%%%%%%%%%%%%%%%%%%%%%%%%%%%%%%%%%%%%%%
\subsection{Multiple inhomogeneities} \label{subsec:multiple}
%%%%%%%%%%%%%%%%%%%%%%%%%%%%%%%%%%%%%%%%%%%

In  the early-universe hadronic gas there is no reason to think that only one bubble of different temperature would form (as opposed to say  a nuclear collision which is a system of very limited  size).  In the absence of data all we can give is an upper bound to the entropy 
produced by disposing as
 many inhomogeneities as possible 
(as long as the background does not lose its meaning). 
We adopt as an extreme limit the density of bubbles when their Gaussian two-sigma walls touch. Thus, we will consider for geometric simplicity a Cartesian arrangement featuring inhomogeneities disposed
as in a simple centered cubic structure. The typical size of each inhomogeneity will be $\approx 2R$ in diameter. We take $4R$ as reasonable average separation between inhomogeneities. The edge of such cube 
has a length of $2R\,N$ due to the presence of $N$ inhomogeneities plus $(N-1)4\,R$ due to the $(N-1)$ spacings, 
 as we show in figure \ref{figure:node}.
\begin{center}
	\begin{figure}
\begin{tikzpicture}
%%%%%%%%%%%%%%%%%%%%%%%%%%%%%%%%%%%%%
% circles
	\draw [very thick] (-4,0) circle (0.4);
	\draw [very thick] (-1.9,0) circle (0.4);
	\draw [very thick] (1.5,0) circle (0.4);	
	\draw [very thick] (3.6,0) circle (0.4);	
%%%%%%%%%%%%%%%%%%%%%%%%%%%%%%%%%%%%%%
% lines 	
	\draw [thick] (-3.6,0) -- (-2.3,0);
	\draw [thick] (-1.5,0) -- (-0.5,0);
	\draw [thick] (0.1,0) -- (1.1,0);
	\draw [thick] (1.9,0) -- (3.2,0);
%%%%%%%%%%%%%%%%%%%%%%%%%%%%%%%%%%%%%
% braces
\draw[decorate,decoration={brace,amplitude=3pt,mirror}] 
(-4.4,-0.6) node(t_k_unten){} -- 
(-3.6,-0.6) node(t_k_opt_unten){}; 
\draw[decorate,decoration={brace,amplitude=3pt,mirror}] 
(-3.5,-0.6) node(t_k_unten){} -- 
(-2.4,-0.6) node(t_k_opt_unten){}; 
\draw[decorate,decoration={brace,amplitude=3pt,mirror}] 
(-2.3,-0.6) node(t_k_unten){} -- 
(-1.5,-0.6) node(t_k_opt_unten){}; 
\draw[decorate,decoration={brace,amplitude=3pt,mirror}] 
(-1.4,-0.6) node(t_k_unten){} -- 
(-0.5,-0.6) node(t_k_opt_unten){}; 
\draw[decorate,decoration={brace,amplitude=3pt,mirror}] 
(0.1,-0.6) node(t_k_unten){} -- 
(1,-0.6) node(t_k_opt_unten){}; 
\draw[decorate,decoration={brace,amplitude=3pt,mirror}] 
(1.1,-0.6) node(t_k_unten){} -- 
(1.9,-0.6) node(t_k_opt_unten){}; 
\draw[decorate,decoration={brace,amplitude=3pt,mirror}] 
(2,-0.6) node(t_k_unten){} -- 
(3.1,-0.6) node(t_k_opt_unten){}; 
\draw[decorate,decoration={brace,amplitude=3pt,mirror}] 
(3.2,-0.6) node(t_k_unten){} -- 
(4,-0.6) node(t_k_opt_unten){}; 
%	\draw (-2,-2) -- (-2,1.2);
%	\draw (0,-1.2) -- (0,2);
%	\filldraw [gray] (-1,-0) circle (2pt);
%	\draw (-1,0) to [out=90,in=190] (2.4,1.6);
%	\draw (-1,0) to [out=-90,in=180] (2.4,-1.6);
%%%%%%%%%%%%%%%%%%%%%%%%%%%%%%%%%%%%%
% sizes
	\draw  (-4,-1.2) node{$2R$};
	\draw  (-1.9,-1.2) node{$2R$};
	\draw  (1.5,-1.2) node{$2R$};
	\draw  (3.6,-1.2) node{$2R$};
	
	\draw  (-2.95,-1.2) node{$4R$};
	\draw  (-0.95,-1.2) node{$4R$};
	\draw  (0.55,-1.2) node{$4R$};
	\draw  (2.55,-1.2) node{$4R$};
%%%%%%%%%%%%%%%%%%%%%%%%%%%%%%%%%%%%
% nodes and points
	\draw [thick] (-0.3,0) circle (0.01);
	\draw [thick] (-0.2,0) circle (0.01);
	\draw [thick] (-0.1,0) circle (0.01);
	\draw  (3.5,0.8) node{$N$};
	\draw  (1.45,0.8) node{$N-1$};
%	\draw (-3,-0) node{$\theta\rightarrow \theta +2\pi$};
%	\draw (2.5,0) ellipse (1 and 1.6);
%	\draw[ ->] (3.8,0.8) arc (0:60:1);
%	\draw[ ->] (-0.5,0) arc (10:60:1);
\end{tikzpicture}

\caption{Sketch of the inhomogeneities arrangement.}\label{figure:node}
\end{figure}
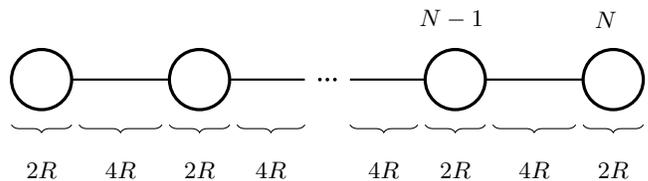
\end{center}
\vspace{-8mm}

The background entropy ${S}^{(N)}_{back}$ for $N$ inhomogeneities occupying a volume $V_C$ is then
\be 
S^{(N)}_{back}=V_c s_{back} \,,
\ee
with $V_c=\left[(N-1)4R +2NR\right]^3$.

Next we need to model the intensity of each perturbation, $\delta T_0$. In a plasma this will be randomly distributed. Conceivable noise models are white noise (all $\delta T_0$ equally likely) or Brownian noise (the distribution falls as $1/\delta T_0^2)$. An interesting intermediate case that is ubiquitous in physics is the so called $1/f$ noise~\cite{noise} that distributes the bubbles in proportion to $1/\delta T_0$.
Both $1/f$ and $1/f^2$ noises obviously assign lower density to higher $\delta T_0$ We currently have no reason to prefer one or another distribution, so we examine all three of them.
In future work we will examine acoustic oscillations of the gas performing a spectral analysis so that the coefficient amplitudes of each Fourier mode will be left arbitrary, to improve the treatment here).

\begin{figure}
\includegraphics[scale=0.67]{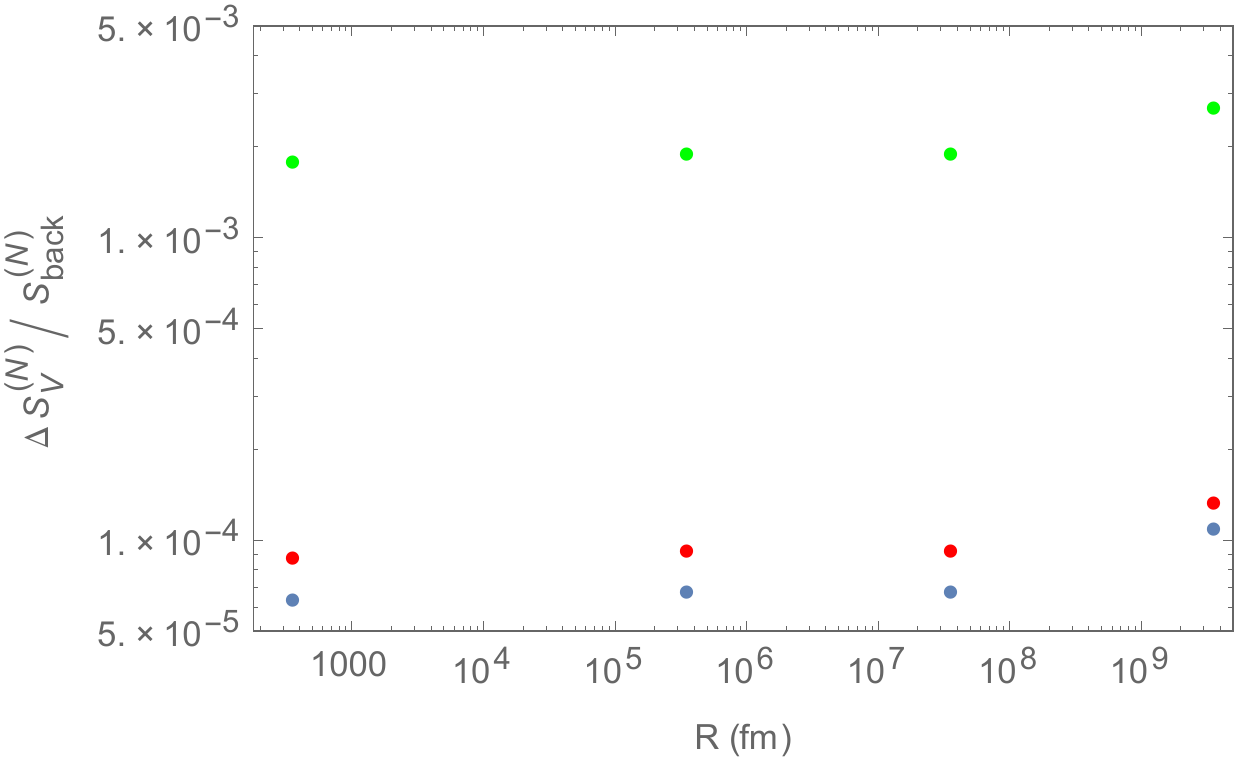}
\caption{$\Delta S^{(N)}_V/S^{(N)}_{back}$ against $R$ at fixed $T_{back}=100$ MeV and $\delta T_0 = 40\,\text{MeV}$. From bottom to top: $1/f^2$ noise function (blue online), $1/f$ noise (red online), constant white noise (green online). \label{figure:entsize}}
\end{figure}

The noise function is in all three cases of the form
\be 
P(\delta T_0)=\frac{\mathcal{C}}{\delta T_0^\beta}\,,
\ee
wherein $\beta=0,1,2$ for the white, $1/f$ and $1/f^2$ noises respectively. The normalization constant $\mathcal{C}$ is determined from the total number of inhomogeneities $N$ in $V_C$ by 
\be 
\int^{\delta T_b}_{\delta T_a} d(\delta T_0)\,\frac{\mathcal{C}}{ \delta T_0^\beta} = N\,,
\ee
with   $\delta T_a\,,\delta T_b$ the lower and upper limits respectively for the initial temperature of the inhomogeneity, i.e., $\delta T_0\in [\delta T_a,\delta T_b]$. 
Too high initial temperatures will involve the quark and gluon plasma and are thus out of our reach here, so $\delta T_b\sim 40$ MeV seems reasonable for this exploration.
As for the smallest $\delta T_0$ taken, since we work in the isospin limit (for example in the computation of the thermal conductivity), it doesn't make sense to retain scales smaller than about $5$ MeV where quark-mass or electromagnetic isospin breaking  effects may play a role.
Thus, we will choose a temperature interval between $40$ MeV to $5$ MeV for the separation above the thermal background. 
 In figure \ref{figure:entsize} we plot $\Delta S^{(N)}_V/S^{(N)}_{back}$ 
for different initial sizes at fixed $T_{back}=100$ MeV for the three noise profiles.

\begin{figure}
\includegraphics[scale=0.65]{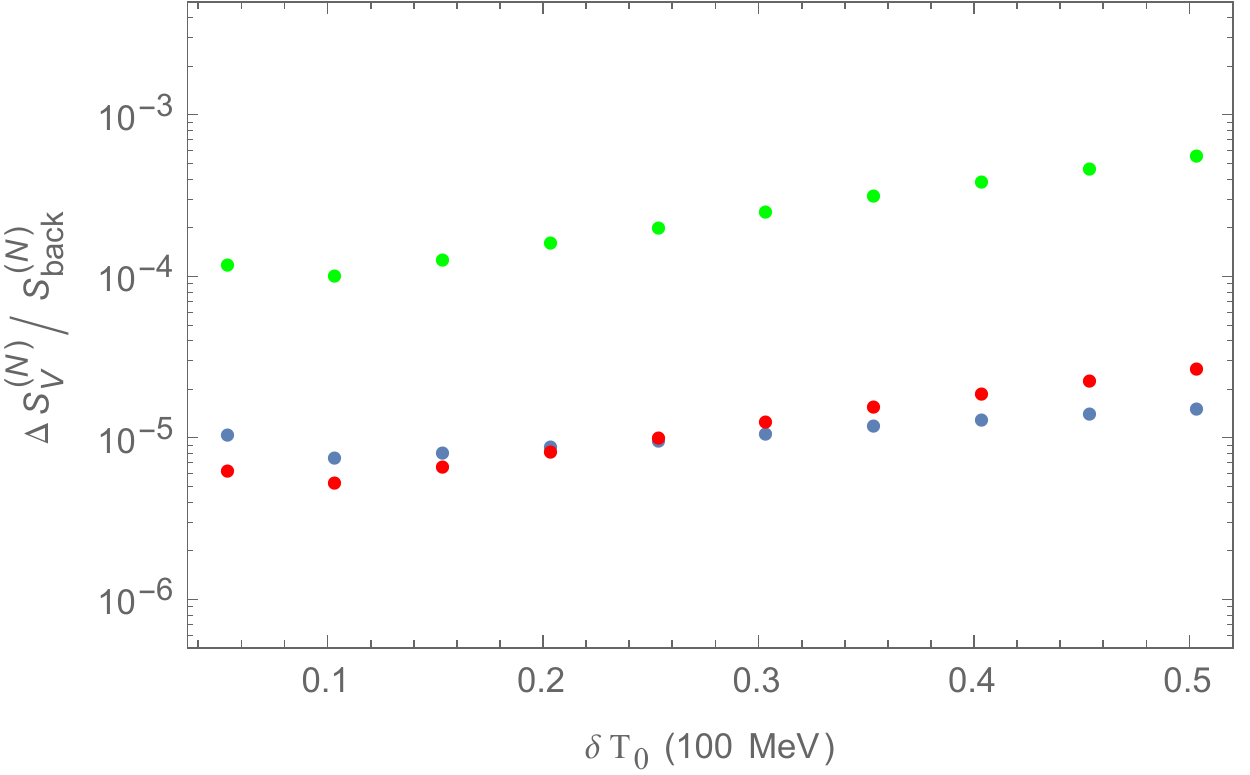}
\caption{$\Delta S^{(N)}_V/S^{(N)}_{back}$ against $\delta T_0$ at $R= 2.5\times 10^{5}\,\text{fm}$ and $T_{back}=100\,\text{MeV}$. Blue online: $1/f^2$ noise function, red online: $1/f$ noise, green online: constant white noise. \label{figure:enttinhom} }
\end{figure}

The summed, integrated entropy $\Delta S^{(N)}_V$ is defined as the integrated entropy summed over all inhomogeneities weighted by function picked, namely,
\be 
\Delta S^{(N)}_V(\delta T_0)=\sum_{\delta T_0}\Delta S_V(\delta T_0)P(\delta T_0)\,,
\ee
with $\Delta S_V$ defined in Eq.~(\ref{intentropy}). Tables \ref{table:entropymultiplef2},\ref{table:entropymultiplef}, and \ref{table:entropymultipleone} give the summed, integrated entropy with distribution of inhomogeneities following the noise functions $1/f^2\,,1/f\,,1$ respectively. In the case of the $1/f\,,1/f^2$ noises, there is a balance between the larger entropy production in the hotter bubbles 
and the larger probability of finding the colder ones, yielding a relatively flat entropy-production dependence on $\delta T_0$. As it is shown in figure \ref{figure:enttinhom}, the largest entropy production is attained for the white noise distribution. Nonetheless, note that in no case $\Delta S^{(N)}_V$ is much greater than about $10^{-6} S^{(N)}_{back}$.\\

\begin{table}[htbp]
\caption{$\Delta S^{(N)}_V/S^{(N)}_{back}$ (in units of $10^{-6}$) for different temperatures. The four columns of data correspond to $R_1=2.5\times 10^{9}\,,R_2=2.5\times 10^{7}\,,R_3=2.5\times 10^5\,,R_4=2.5\times 10^2$ 
in fm units. $T_{back}$ and $\delta T_0$ are expressed in MeV.
\label{table:entropymultiplef2}}
\begin{tabular}{|c @ {\hspace{2mm}} c @ {\hspace{5mm}} |c @ {\hspace{5mm}} c @ {\hspace{5mm}} c @ {\hspace{5mm}} c @ {\hspace{0mm}} |}
\toprule
\multicolumn{2}{|c|}{ $ $ } &
\multicolumn{4}{c|}{ Noise function $1/f^2$}\\
$T_{back}$ & $\delta T_0$ &  1 & 2& 3 & 4\\
\toprule
\multirow{8}{0.55cm}{120}
 & 40 & 8.1 & 5.3 & 5.3 & 5.0\\
 & 35 & 7.2 & 4.5 & 4.5 & 4.2 \\
 & 30 & 6.6 & 4.1 & 4.0 & 3.4 \\
 & 25 & 6.3 & 2.9 & 2.9 & 2.6  \\
 & 20 & 4.3  & 2.5 & 2.5 & 2.4 \\
 & 15 & 4.0 & 1.5  & 1.5  & 1.4 \\
 & 10 & 6.2 & 9.8 $\times 10^{-1}$ & 9.8 $\times 10^{-1}$ & 9.1 $\times 10^{-1}$ \\
  & 5 & 7.2 & 2.5 $\times 10^{-1}$ & 3.1 $\times 10^{-1}$ & 1.4 $\times 10^{-1}$ \\
\hline
\multirow{8}{0.55cm}{100}

 & 40 & 13.0 & 10.6 & 10.6 & 10.0 \\
  & 35 & 11.9 & 9.0 & 9.0 & 8.5 \\
 & 30 &  10.6 & 7.5 & 7.5 & 7.0 \\
 & 25 &  9.6 & 5.8 & 5.8 & 5.5  \\
 & 20 & 8.9 & 4.3 & 4.3 & 4.0 \\
 & 15 &  8.1 & 2.7 & 2.7 & 2.6 \\
 & 10 & 7.6 & 1.4 & 1.4 & 1.3 \\
  & 5 & 10.5 & 4.1 $\times 10^{-1}$ & 4.1 $\times 10^{-1}$ & 3.9 $\times 10^{-1}$ \\
\toprule
\end{tabular}
\end{table}

\begin{table}[htbp]
\caption{$\Delta S^{(N)}_V/S^{(N)}_{back}$ (in units of $10^{-6}$) for different temperatures. The four columns of data correspond to $R_1=2.5\times 10^{9}\,,R_2=2.5\times 10^{7}\,,R_3=2.5\times 10^5\,,R_4=2.5\times 10^2$ in fm units. $T_{back}$ and $\delta T_0$ are given in MeV.
\label{table:entropymultiplef}}
\begin{tabular}{|c @ {\hspace{2mm}} c @ {\hspace{5mm}}| c @ {\hspace{5mm}} c @ {\hspace{5mm}} c @ {\hspace{5mm}} c @ {\hspace{0mm}} |}
\toprule
\multicolumn{2}{|c|}{ $ $ } &
\multicolumn{4}{c|}{ Noise function $1/f$}\\
$T_{back}$ & $\delta T_0$ &1 &  2 &3 & 4\\
\toprule

\multirow{8}{0.55cm}{120}
& 40 & 9.8 & 7.8  & 7.8  & 7.0  \\
& 35 & 7.9 & 5.9  & 5.9  & 4.8  \\
& 30 & 6.2 & 4.4 & 4.4 & 3.8  \\
& 25 & 4.9 & 3.6 & 3.6 & 2.2  \\
& 20 & 4.6 & 1.8 & 1.8 & 1.6  \\
& 15 & 5.5 &  1.1  &  1.1 & 9.4$\times 10^{-1}$\\
& 10 & 4.6 & 7.2$\times 10^{-1}$ & 7.2$\times 10^{-1}$ & 5.3$\times 10^{-1}$\\
& 5 & 5.7 & 1.8$\times 10^{-1}$ & 1.8$\times 10^{-1}$ & 1.4$\times 10^{-1}$\\

\hline
\multirow{8}{0.55cm}{100}

& 40 & 18.9 & 15.4 & 15.4 & 14.5\\
& 35 & 15.6 & 11.9 & 11.9 & 11.2\\
& 30 & 12.6 & 8.8 & 8.8 & 8.3\\
& 25 & 10.0 & 6.1 & 6.1 & 5.8\\
& 20 & 8.2 & 3.9 & 3.9 & 3.7\\
& 15 & 6.6 & 2.2 & 2.2 & 2.1\\
& 10 & 5.3 & 9.9$\times 10^{-1}$ & 9.9$\times 10^{-1}$ & 9.4$\times 10^{-1}$\\
& 5 & 6.3 & 2.5$\times 10^{-1}$ & 2.5$\times 10^{-1}$ & 2.3$\times 10^{-1}$\\

\toprule

\end{tabular}
\end{table}

\begin{table}[htbp]
\caption{$\Delta S^{(N)}_V/S^{(N)}_{back}$ (in units of $10^{-6}$) for different temperatures. The four columns of data correspond to $R_1=2.5\times 10^{9}\,,R_2=2.5\times 10^{7}\,,R_3=2.5\times 10^5\,,R_4=2.5\times 10^2$ in fm units. $T_{back}$ and $\delta T_0$ are both given in MeV. 
\label{table:entropymultipleone}}
\begin{tabular}{|c @ {\hspace{2mm}} c @ {\hspace{5mm}} |c @ {\hspace{5mm}} c @ {\hspace{5mm}} c @ {\hspace{5mm}} c @ {\hspace{0mm}} |}
\toprule
\multicolumn{2}{|c|}{ $ $ } &
\multicolumn{4}{c|}{ White noise}\\
$T_{back}$ & $\delta T_0$ & 1 &  2 &  3& 4\\
\toprule

\multirow{8}{0.55cm}{120}

& 40 & 195.1 & 166.3 & 166.3 & 149.8  \\
& 35 & 195.5 & 120.1 & 120.1 & 112.6 \\
& 30 &  183.4 & 120.6 & 120.5 & 107.1\\
& 25 & 153.6 & 84.7 & 84.7 & 78.7 \\
& 20 & 137.3 & 64.3 & 64.3 & 50.4\\
& 15 &  126.1 & 30.4 & 30.4 & 24.3\\
& 10 &  57.3 & 12.8 & 12.82 & 12.3\\
& 5 & 52.2 & 2.7 & 2.7 & 2.6 \\
\hline

\multirow{8}{0.55cm}{100} 

& 40 & 386.5 & 315.9 & 315.8 & 297.5 \\
& 35 & 315.7 & 240.7 & 240.7 & 226.9 \\
& 30 & 251.5 & 175.9 & 175.9 & 165.9 \\
& 25 & 199.0 & 121.3 & 121.3 & 114.5 \\
& 20 & 160.6 & 77.1 & 77.0 & 72.8 \\
& 15 & 127.5 & 43.0 & 42.9 & 40.6 \\
& 10 & 101.1 & 18.9 & 18.90 & 17.9 \\
& 5 & 118.6 & 4.7 & 4.6 & 4.4 \\

\toprule

\end{tabular}
\end{table}

The outcome of the computation is that the entropy produced is only a small fraction of the background entropy in the same volume, because many of the inhomogeneities are just of small intensity with these weight functions. But these are ad hoc: independent ways of assessing what inhomogeneities are possible need to be found and we look forward to progress in that respect. 

%%%%%%%%%%%%%%%%%%%%%%%%%%%%%%%%%%%%%%%%%%%%%%%%%%%%%%%%%%%%%%%%%%%%%%%%%%%%%%
\section{Conclusions}
%%%%%%%%%%%%%%%%%%%%%%%%%%%%%%%%%%%%%%%%%%%%%%%%%%%%%%%%%%%%%%%%%%%%%%%%%%%%%%

In this work we have examined the production of entropy by thermal inhomogeneities in the pion gas produced after the quark-gluon plasma hadronization at the early universe.

In view of the uniformity of the CMB at large scales, too large to not have been in causal contact, standard theory 
invokes a time of accelerated expansion (inflation) of a universe in thermal equilibrium. Thus, it would appear natural to assume that such equilibrium was also reached early-on at small scales. Nevertheless, the naturality argument is not fool proof (recall the recent discovery of a light Higgs with no accompanying supersymmetric partners for any of the SM particles) and does not discard the possible existence of small-scale inhomogeneities in the early universe. If any such did not have enough time to dissipate before the quark-gluon plasma decay cross-over, or was produced during that phase transition,
the entropy production is calculable with modern nuclear and particle physics theory. We have exemplified with the computation of such entropy increase in a thermal inhomogeneity in the pion gas, but the field is ample and much more work is possible. 

We have concentrated on the pion gas at a time interval when pions were 
main contributors to the universe's entropy, even more than photons,
and the ones carrying the largest possible inhomogeneity due to their short mean free path.

Of course, further processes involving non-vanishing entropy production, such as flavor or momentum diffusion, are expected to be of equal potential importance. We leave them for future work. Here we have remained within the realm of small, Newtonian thermal perturbations, and the produced $\Delta S_V/S_{\rm back}$ is thus a small fraction.
In future work we plan to examine the damping of acoustic oscillations in this phase by examining subJeans modes in the presence of dissipation coefficients, and try to put quantitative constraints on the maximum size of inhomogeneities that can be dissipated. Looking at figure~\ref{figure:deltat}, we see that the pion gas can easily reduce a thermal perturbation at the 30\% level down to the 3\% level, that is, an order of magnitude. A more detailed study beyond this first exploration is granted.
Meanwhile, for inhomogeneities of that intensity, production of entropy is significant, as can be seen in the various tables of subsection~\ref{subsec:multiple}.

To conclude, we have pointed out that entropy production deserves being examined in the hadron-lepton phase between 1 and 175 MeV, and we have studied in detail one example, that of relaxation of thermal inhomogeneities in the pion gas.

%%%%%%%%%%%%%%%%%%%%%%%%%%%%%%%%%%%%%%%%%%%%%%%%%%%%%%%%%%%%%%%%%%%%%%%%%%%%%%
\section*{Acknowledgments}
%%%%%%%%%%%%%%%%%%%%%%%%%%%%%%%%%%%%%%%%%%%%%%%%%%%%%%%%%%%%%%%%%%%%%%%%%%%%%%
We thank Antonio Maroto for a critical reading of the cosmology aspects of the work.
Supported by the Spanish Excellence Network on Hadronic Physics FIS2014-57026-REDT, and by grants UCM:910309, MINECO:FPA2011-27853-C02-01, MINECO:FPA2014-53375-C2-1-P and CPAN Consolider-Ingenio 2010. DRF was partially supported by a GRUPIN 14-108 research grant from Principado de Asturias.

%%%%%%%%%%%%%%%%%%%%%%%%%%%%%%%%%%%%%%%%%%%% 
 
%%%%%%%%%%%%%%%%%%%%%%%%%%%%%%%%%%%%%%%%%%%% 


\begin{thebibliography}{99} 
%%%%%%%%%%%%%%%%%%%%%%%%%%%%%%%%%%%%%%%%%%%% 




%\cite{Rafelski:2013yka}
\bibitem{Rafelski:2013yka} 
  But see J.~Rafelski and J.~Birrell,
  %``Traveling Through the Universe: Back in Time to the Quark-Gluon Plasma Era,''
  J.\ Phys.\ Conf.\ Ser.\  {\bf 509}, 012014 (2014);
%  [arXiv:1311.0075 [nucl-th]]
%\cite{Birrell:2014cja}
%\bibitem{Birrell:2014cja} 
%  J.~Birrell and J.~Rafelski,
  %``Connection of Cosmic Microwave Background Fluctuations to the Quark-Gluon Hadronization Temperature,''
  {\it ibid.} arXiv:1404.6005 [nucl-th].
  %%CITATION = ARXIV:1404.6005;%%



%\cite{Rafelski:2013qeu}
\bibitem{Rafelski:2013qeu} 
  J.~Rafelski,
  %``Connecting QGP-Heavy Ion Physics to the Early Universe,''
  Nucl.\ Phys.\ Proc.\ Suppl.\  {\bf 243-244}, 155 (2013);
  %%CITATION = ARXIV:1306.2471;%%
%\cite{Birrell:2014uka}
%\bibitem{Birrell:2014uka} 
  J.~Birrell, C.~-T.~Yang and J.~Rafelski,
  %``Relic Neutrino Freeze-out: Dependence on Natural Constants,''
  arXiv:1406.1759 [nucl-th];
 A. Faessler {\it et al.} EPJ Web of Conf. {\bf 71}, 00044 (2014).


%\cite{Brambilla:2014jmp}
\bibitem{Brambilla:2014jmp} See the review in pages 111-112 of
  N.~Brambilla  {\it et al.},
   Eur.\ Phys.\ J.\  {\bf C74},  (2014) 10, 2981;
  %``QCD and Strongly Coupled Gauge Theories: Challenges and Perspectives,''
  arXiv:1404.3723 [hep-ph].
  %%CITATION = ARXIV:1404.3723;%%

%\cite{Torres-Rincon:2012sda}
\bibitem{Torres-Rincon:2012sda} 
  J.~M.~Torres-Rincon,   10.1007/978-3-319-00425-9
  ``Hadronic Transport Coefficients from Effective Field Theories,''
  Dissertation presented to the Univ. of Madrid (Complutense) available as a Springer thesis 2013,
DOI: 10.1007/978-3-319-00425-9,  arXiv:1205.0782 [hep-ph]. Older results on the thermal conductivity and viscosity may be found in
  %%CITATION = ARXIV:1205.0782;%%
  A.~Dobado, F.~J.~Llanes-Estrada and J.~M.~Torres Rincon, Proceedings of the IVth International Conference on Quarks and Nuclear Physics, Madrid, 2006,
  %``Heat conductivity of a pion gas,''
  hep-ph/0702130 [HEP-PH];
  A.~Dobado and F.~J.~Llanes-Estrada,
  %``The Viscosity of meson matter,''
  Phys.\ Rev.\ D {\bf 69}, 116004 (2004).
  %%CITATION = HEP-PH/0309324;%%
  %%CITATION = HEP-PH/0702130;%%
Diffusion of heavy flavors is treated in
  L.~M.~Abreu {\it et al.}
  %``Charm diffusion in a pion gas implementing unitarity, chiral and heavy quark symmetries,''
  Annals Phys.\  {\bf 326}, 2737 (2011);
  %%CITATION = ARXIV:1104.3815;%%
  D.~Cabrera {\it et al.}
  %``Strange and heavy mesons in hadronic matter,''
  J.\ Phys.\ Conf.\ Ser.\  {\bf 503}, 012017 (2014);
  %%CITATION = ARXIV:1312.4343;%%
the electric conductivity can be found in
 D.~Fernandez-Fraile and A.~Gomez Nicola,
  %``The Electrical conductivity of a pion gas,''
  Phys.\ Rev.\ D {\bf 73}, 045025 (2006).
  %%CITATION = HEP-PH/0512283;%%



%\cite{Davesne:1995ms}
\bibitem{Davesne:1995ms} 
  D.~Davesne,
  %``Transport coefficients of a hot pion gas,''
  Phys.\ Rev.\ C {\bf 53}, 3069 (1996).
  %%CITATION = PHRVA,C53,3069;%%

%\cite{Prakash:1993kd}
\bibitem{Prakash:1993kd} 
  M.~Prakash, M.~Prakash, R.~Venugopalan and G.~M.~Welke,
  %``How fast is equilibration in hot hadronic matter?,''
  Phys.\ Rev.\ Lett.\  {\bf 70}, 1228 (1993)
  [Nucl.\ Phys.\ A {\bf 566}, 403C (1994)].
  %%CITATION = PRLTA,70,1228;%%

%\cite{Mitra:2014dia}
\bibitem{Mitra:2014dia} 
  S.~Mitra and S.~Sarkar,
  %``Medium effects on the thermal conductivity of a hot pion gas,''
  Phys.\ Rev.\ D {\bf 89}, no. 5, 054013 (2014)
  [arXiv:1403.3554 [nucl-th]].
  %%CITATION = ARXIV:1403.3554;%%

%\cite{Labini:2011tj}
\bibitem{Labini:2011tj} 
  F.~S.~Labini,
  %``Inhomogeneities in the universe,''
  Class.\ Quant.\ Grav.\  {\bf 28}, 164003 (2011).
  %%CITATION = ARXIV:1103.5974;%%

%\cite{Abelev:2013haa}
\bibitem{Abelev:2013haa} 
  B.~B.~Abelev {\it et al.}  [ALICE Collaboration],
  %``Multiplicity Dependence of Pion, Kaon, Proton and Lambda Production in p-Pb Collisions at $\sqrt{s_{NN}}$ = 5.02 TeV,''
  Phys.\ Lett.\ B {\bf 728}, 25 (2014)
  [arXiv:1307.6796 [nucl-ex]].
  %%CITATION = ARXIV:1307.6796;%%


\bibitem{Planck} Planck Collaboration (P.A.R. Ade et al.), Planck 2015
results. XIII. Cosmological parameters, arXiv:1502.01589
[astro-ph.CO]. P. A. R. Ade et al. [Planck Collaboration],
Planck 2015. XX. Constraints on inflation,
arXiv:1502.02114 [astro-ph.CO].

\bibitem{part} 
J.~Beringer {\it et al.}  [Particle Data Group Collaboration],
%``Review of Particle Physics (RPP),''
Phys.\ Rev.\ D {\bf 86}, 010001 (2012).
%%CITATION = PHRVA,D86,010001;%%

%\cite{Burles:2000zk}
\bibitem{Burles:2000zk} 
  S.~Burles, K.~M.~Nollett and M.~S.~Turner,
  %``Big bang nucleosynthesis predictions for precision cosmology,''
  Astrophys.\ J.\  {\bf 552}, L1 (2001)
  [astro-ph/0010171].
  %%CITATION = ASTRO-PH/0010171;%%

\bibitem{Weinberg} \emph{Cosmology}, Steven Weinberg, $1^{st}$ edition, 2008 Oxford University Press.


%\bibitem{term} \emph{Understanding non-equilibrium thermodynamics} G. Lebon, D. Jou, J. Casas-V\'azquez,  Springer-Verlag (Heidelberg-Berlin) $1^{st}$ edition, 2008.


\bibitem{noise} E.~Milotti, 
Invited paper at the ``2ndo. Encuentro del Grupo Latinoamericano de Emision Acustica y 1ro. Iberoamericano, E-GLEA-2'', Buenos Aires (Argentina), 11-14 september 2001
arXiv:physics/0204033.



%%%%%%%%%%%%%%%%%%%%%%%%%%%%%%%%%%%%%%%%%%%% 
\end{thebibliography}
\end{document}